\documentclass[prl,superscriptaddress,floatfix,showpacs,aps,reprint,letterpaper,nobalancelastpage]{revtex4-1}
\usepackage{graphicx} 
\usepackage{color} 
\usepackage{amsmath}
\usepackage{hyperref} 
\usepackage{amssymb}

\newcommand{\ket}[1]{\left\lvert #1 \right\rangle}

\newcommand{\MHz}{\mathrm{MHz}}
\newcommand{\GHz}{\mathrm{GHz}}
\newcommand{\us}{\mu\mathrm{s}}
\newcommand{\ns}{\mathrm{ns}}

\newcommand{\tg}{t_{\text{g}}}

\newcommand{\Jeff}{J_{\text{eff}}}
\newcommand{\CR}{\text{CR}_{12}}

\begin{document}
\title{A simple all-microwave entangling gate for fixed-frequency superconducting qubits}
\date{June 2, 2011}

\author{Jerry M. Chow}
\affiliation{IBM T.J. Watson Research Center, Yorktown Heights, NY 10598, USA}
\author{A. D. C\'orcoles}
\affiliation{IBM T.J. Watson Research Center, Yorktown Heights, NY 10598, USA}
\author{Jay M. Gambetta}
\affiliation{IBM T.J. Watson Research Center, Yorktown Heights, NY 10598, USA}
\author{Chad Rigetti}
\affiliation{IBM T.J. Watson Research Center, Yorktown Heights, NY 10598, USA}
\author{B. R. Johnson}
\affiliation{Raytheon BBN Technologies, Cambridge, MA 02138, USA}
\author{John A. Smolin}
\affiliation{IBM T.J. Watson Research Center, Yorktown Heights, NY 10598, USA}
\author{J. R. Rozen}
\affiliation{IBM T.J. Watson Research Center, Yorktown Heights, NY 10598, USA}
\author{George A. Keefe}
\affiliation{IBM T.J. Watson Research Center, Yorktown Heights, NY 10598, USA}
\author{Mary B. Rothwell}
\affiliation{IBM T.J. Watson Research Center, Yorktown Heights, NY 10598, USA}
\author{Mark B. Ketchen}
\affiliation{IBM T.J. Watson Research Center, Yorktown Heights, NY 10598, USA}
\author{M. Steffen}
\affiliation{IBM T.J. Watson Research Center, Yorktown Heights, NY 10598, USA}

\begin{abstract}
We demonstrate an all-microwave two-qubit gate on superconducting qubits which are fixed in frequency at optimal bias points. The gate requires no additional subcircuitry and is tunable via the amplitude of microwave irradiation on one qubit at the transition frequency of the other. We use the gate to generate entangled states with a maximal extracted concurrence of $0.88$ and quantum process tomography reveals a gate fidelity of $81\%$.
\end{abstract}
\maketitle

A basic requirement for fault tolerant quantum computing is a universal set of nearly perfect one- and two-qubit gates. As high-fidelity single-qubit operations on superconducting qubits become routine~\cite{lucero_gates_2008,Chow2010a}, the focus shifts onto developing robust and scalable two-qubit gates. Already, rapid progress has been made, including a controlled-NOT (CNOT) gate with fixed coupled qubits~\cite{Plantenberg:2007it}, and highly entangled states of two~\cite{dicarlo_2009,Ansmann:2009yq} and three~\cite{DiCarlo2010,Neeley2010} qubits generated from tuning qubits to explicit resonances.

Although scaling up superconducting systems with many fixed mutual couplings between qubits is simple to experimentally design, it becomes difficult to control the effective interaction between qubits. Alternatively, this control can be achieved by 1)~tuning the coupling energy between the qubits; or 2)~dynamically changing the detuning between qubits in the presence of some small fixed coupling. In the first case, the coupling takes the form of a non-linear tunable subcircuit which can be driven with either microwaves \cite{Bertet2006,Niskanen:2007if,Harris2007} or dc~\cite{Hime:2006ef,Bialczak2011,Sri2011}. This scheme has the benefit of allowing the qubits to be operated at their optimal bias points for coherence. However, the additional control lines for the tunable subcircuit can also result in added circuit complexity. In the second case, which requires no additional controls other than those for operating the individual qubits, two-qubit gates have been demonstrated such as $\sqrt{i\text{SWAP}}$~\cite{steffen_entang_2006,Bialczak2010} and conditional phase~\cite{dicarlo_2009} by tuning the qubit energy levels into explicit resonance conditions. Although this scheme has been effective for systems up to three qubits~\cite{DiCarlo2010,Neeley2010}, tuning qubit frequencies in devices with even more qubits could lead to unwanted coupling to noncomputational energy levels of the system and to spurious modes of the electromagnetic environment. Hence, desiderata for a scalable qubit coupling would combine tunability of the effective coupling strength with the simplicity of fixed coupling, in an architecture amenable to a larger number of qubits.

In this Letter, we demonstrate a new two-qubit gate which combines the hardware simplicity of a fixed coupling scheme with a tunable effective interaction enacted with only microwave control~\cite{Paraoanu2006,Rigetti2010,Groot2010}, all on a quantum bus architecture~\cite{majer_coupling_2007,Sillanpaa_2007}. Two capacitively-shunted flux qubits~\cite{Steffen2010} (CSFQs) are dispersively coupled through a microwave cavity and parked at locations of optimal coherence. We find that a two-qubit interaction turns on linearly with the amplitude of an applied cross-resonant (CR) drive, in which microwaves resonant with a target qubit are applied on the other control qubit. Up to single-qubit rotations, the CR two-qubit gate is related to the canonical CNOT, and we use it to generate entangled states with a maximal extracted concurrence of $0.88$. Furthermore, quantum process tomography reveals a gate fidelity of $81\%$, with residual errors due to coherence times and single-qubit gate calibration.

\begin{figure}[t!]
\centering
\includegraphics[width=0.45\textwidth]{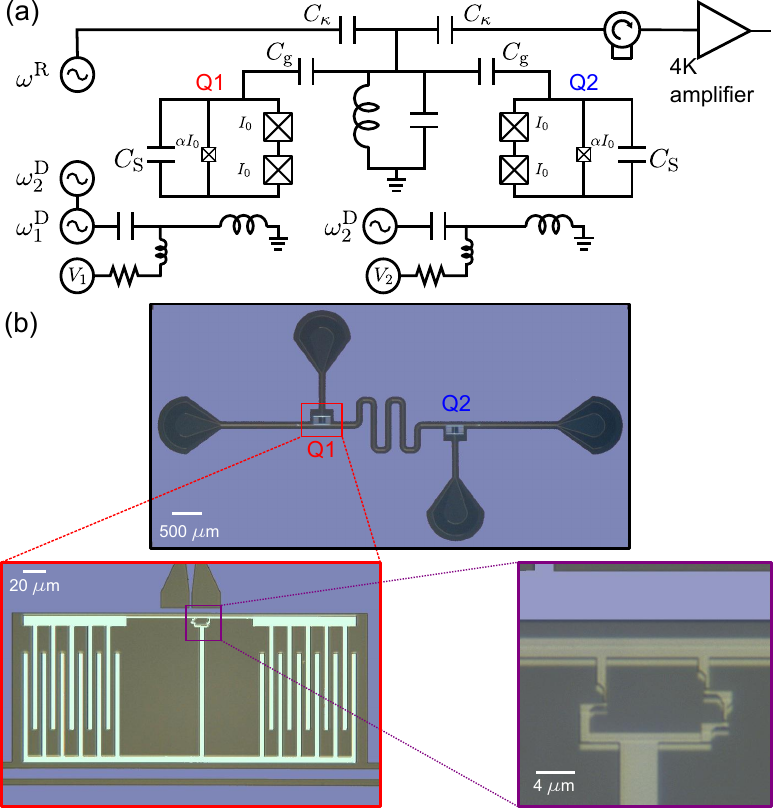}
\caption{\label{fig:1} (color online)~Circuit schematic and two-qubit device.~ (a) Circuit schematic showing two CSFQs (Q1, red on left and Q2, blue on right) with shunt capacitance $C_{\text{S}}$ and small junction ratio $\alpha$, coupled to a single resonator. The qubit-cavity coupling is governed by $C_{\text{g}}$.  Each qubit has an on-chip local flux-bias line which is used to both dc tune the energy levels and serve as a microwave excitation port for driving transitions. Two-qubit joint readout is performed by probing the system through the input port near the cavity frequency and detecting the transmission at the output port. (b) Optical micrographs of device (false-colored). The resonator is realized as a coplanar waveguide with measured frequency $\omega_{\text{R}}/2\pi = 9.72\,\GHz$ and linewidth $\kappa/2\pi = 1\,\MHz$. The flux-bias lines are terminated with an inductance to ground, off-centered from each qubit-loop. Fabrication details are given in previous work~\cite{Steffen2010}}
\end{figure}

CSFQs are a suitable choice for testing the CR protocol, since they have been shown to give consistently long coherence times in a circuit QED scheme~\cite{Steffen2010}. Figures~\ref{fig:1}(a-b) show the schematic of our experimental setup and optical images of our device, in which two qubits are coupled to opposite ends of a coplanar-waveguide resonator ($\omega_{\text{R}}/2\pi = 9.72\,\GHz$) and on-chip flux bias lines (FBLs) are used to independently tune them to their flux sweet-spot transition frequencies,  $\omega_1/2\pi = 5.854\,\GHz$ and $\omega_2/2\pi = 5.528\,\GHz$. Here, we find optimal relaxation [$T_1^{\text{1(2)}}=1.6\,(1.5)\,\us$] and decoherence [$T_2^{*,\text{1(2)}}=1.6\,(1.5)\,\us$] times for both qubits. Operating in the dispersive regime of circuit QED, we measure cavity shifts $\chi_1/\pi = 1.1\,\MHz$ and $\chi_2/\pi = 0.6\,\MHz$ permitting a joint two-qubit readout~\cite{filipp_joint_2009, Chow2010}. To avoid errors due to the finite qubit anharmonicities, $\alpha_1/2\pi = \omega_1^{12}-\omega_1^{01} = 224$ MHz and $\alpha_2/2\pi = \omega_2^{12}-\omega_2^{01} = 255$ MHz, we use gaussians with quadrature derivative pulse-shaping $\sigma = 4\,\ns$, total gate length $4\sigma$,  for single-qubit gates~\cite{Motzoi:2009fx}, $\{X,Y,X_{\pm90},Y_{\pm90}\}$. We use the notation $A_{\theta}=\exp(-i\theta A \pi/360)$ for a rotation of $\theta$ around $A$, and drop the subscript for Pauli operators. The standard deviation of the gaussian shapes is $\sigma = 4\,\ns$ with total gate length $4\sigma$ and the derivative scale parameter~\cite{Chow2010a} is experimentally determined to be $-1.4$ for both qubits.

\begin{figure}[t!]
\centering
\includegraphics[width=0.45\textwidth]{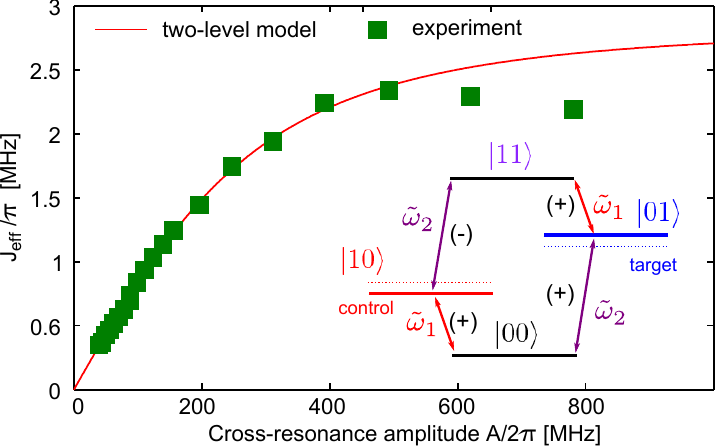}
\caption{ (color online) Cross-resonance level diagram and experimentally extracted tunable coupling strength. The effective interaction strength for different cross-drive powers is found from the extracted frequency shift of the $\Omega_{\text{R},2}$ with qubit 1 in either the ground or excited state. The interaction turns on linearly with the amplitude $A$ of the drive, here shown normalized on the $x$-axis to $\Omega_{\text{R},2}$, before leveling off at higher amplitudes when $\Omega_{\text{R},2}$ approaches $\Delta_{12}$. The maximum interaction strength of $1.4\,\MHz$ is observed at $A/2\pi=493\,\MHz$.(inset) Energy spectrum corresponding to a pair of fixed weakly-coupled qubits ($\Delta_{12}>J$). Dashed (solid) red and blue lines reflect uncoupled (coupled) energy levels for qubit 1 and 2 respectively. Assuming qubit 1 as the control qubit, a cross-drive at the qubit 2 transition frequency rotates qubit 2, the target, with a phase dependent on the state of the control, with a rate down by a factor $J/\Delta_{12}$ over a resonant drive.}
\label{fig:2}
\end{figure}

We implement the CR scheme on our device by applying microwave excitations resonant with the opposite qubit's transition frequency directly onto either qubit via the FBLs [Fig.~\ref{fig:1}(a)]. To understand how the CR effect arises, consider the Hamiltonian for a pair of qubits which are detuned from the resonator by $\Delta_i = \omega_i - \omega_{\text{R}}$ for $i=1,2$, and dispersively coupled to each other via the resonator,
\begin{equation}
	H/\hbar = \frac{1}{2}\omega_1 ZI + \frac{1}{2}\omega_2 IZ +J XX,
	\label{eq:dispH}
\end{equation}
where $\{I,X,Y,Z\}^{\otimes 2}$ are the Pauli operators (including the identity) and the order indexes the qubit number. Equation~\ref{eq:dispH} can be diagonalized and considered as a new set of two qubits with shifted frequencies $\tilde{\omega}_1 = \omega_1 + J/\Delta_{12}$, $\tilde{\omega}_2 = \omega_2 - J/\Delta_{12}$ when $J$ is small compared to the qubit-qubit detuning, $\Delta_{12} = \omega_1-\omega_2$ (see Fig.~\ref{fig:2} inset). In this frame, a single drive on qubit 1 at either $\tilde{\omega}_1$ or $\tilde{\omega}_2$ can excite transitions to qubit 1 or 2, respectively. However, the CR drive amplitude of qubit 2 is reduced by a factor of $J/\Delta_{12}$ and acquires a phase which is dependent on the state of qubit 1. The drive Hamiltonian then takes the form
\begin{equation}
	H_{\text{D}} = \hbar A(t)\cos(\tilde{\omega}_2t)\left(XI-\frac{J}{\Delta_{12}}ZX+m_{12}IX\right),
\label{eq:driveH}
\end{equation}
where $A(t)$ is the shaped microwave amplitude of a drive on qubit 1 and $m_{12}$ represents spurious crosstalk due to stray electromagnetic coupling in the device circuit and package~\cite{Wenner2011}. Hence, a drive on qubit 1 at $\tilde{\omega}_2$ can be used to turn on a $ZX$ interaction, which is a primitive~\cite{Paraoanu2006} for the two-qubit CNOT. The same analysis holds symmetrically for a drive applied to qubit~2. We will use the notation $\text{CR}_{ij}(A,\tg)$ to represent a cross drive on qubit $i$ at $\omega_j$ with amplitude $A$ and gate time $\tg$.

\begin{figure*}[t!]
\centering
\includegraphics[width=0.9\textwidth]{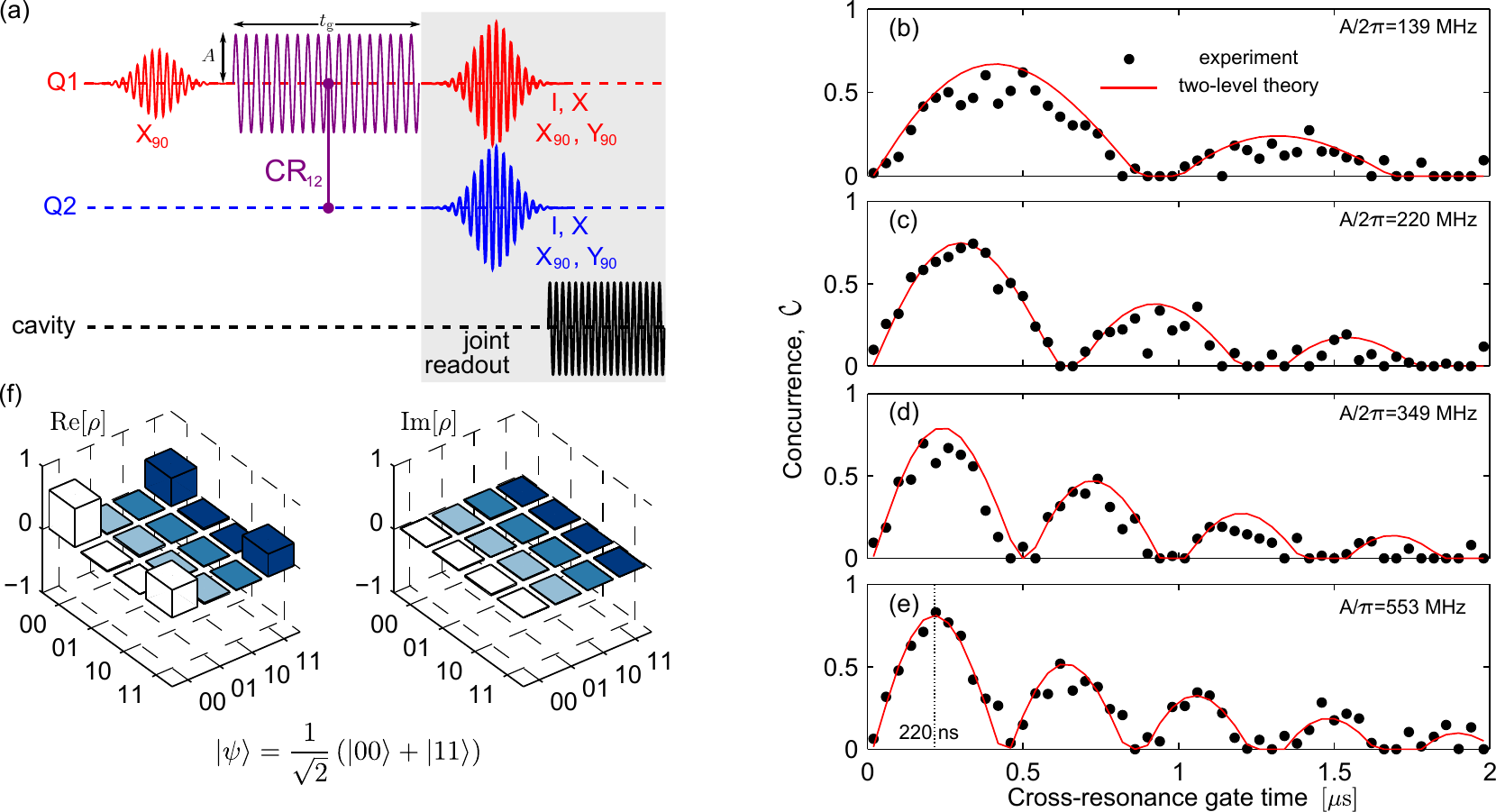}
\caption{(color online)~Entangled states and concurrence oscillations. (a)~Pulse sequence for generating entangled states; both qubits are initialized in the ground state; qubit 1 is first placed into a superposition state with a $X_{+90}$, leaving the system in the separable state $\ket{\psi} = (\ket{00}+\ket{10})/\sqrt{2}$; next the $\CR$ pulse is also applied to qubit 1 before the joint readout sequence. The concurrence can be computed for all density matrices obtained with this pulse protocol, and oscillations are observed as a function of the gate time $\tg$ for four different $\CR$ drive amplitudes (b-e), corresponding to $\{139, 220, 349, 553\}$ MHz. The period of the oscillations correspond to $1/\Jeff$ and the maximum concurrence is observed at $\tg= 220$ ns at $A/2\pi=553\,\MHz$. (f) Measured density matrix for Bell-state $\ket{\psi_{\text{Bell}}} = (\ket{00} +\ket{11})/\sqrt{2}$ generated at the point of optimal concurrence labeled in (e).}
\label{fig:3}
\end{figure*}

Although a $ZX$ interaction theoretically corresponds to an $X$ rotation on qubit 2 with the direction dependent on the state of qubit 1, in practice due to the $m_{12}\sim 0.5$ term in Eq.~\ref{eq:driveH}, $\CR$ also directly induces an additional rotation of qubit 2. This spurious crosstalk parameter $m_{12}$ is determined by comparing Rabi frequencies of both qubits when driven with the same amplitude through the same FBL. This effect does not degrade the two-qubit interaction because it commutes with the $ZX$ term. The effective interaction strength $\Jeff$ is then manifested as the difference in qubit 2 Rabi oscillation frequencies, $\Omega_{\text{R},2}$, dependent on the state of qubit 1.

Figure~\ref{fig:2}(b) shows the experimentally measured $\Jeff$ versus $A$. We shape the $\CR$ pulse as a slow gaussian turn-on with a flat-top and a derivative-pulse correction on the quadrature (scale parameter of 0.8). With and without a single qubit $X$ gate on qubit 1, we find different $\Omega_{\text{R},2}$, extracted from oscillations of the qubit 2 excited state population versus the time of the $\CR$ pulse $\tg$. For small $A$, $\Jeff$ turns on linearly. However, at stronger drives, the interaction strength levels off to a maximum of $1.4\,\MHz$, which is in agreement with a two-level theory~\cite{Rigetti2010} and is due to the off-resonant driving of $XI$ in eq.~\ref{eq:driveH}. At the strongest of drives the measured $\Jeff$ does not agree with the two-level theory due to the presence of higher levels in the qubits and the breakdown of our simplified derivative-pulse shaping correction.

Non-classical states can be generated and measured using the protocol in figure~\ref{fig:3}a, in which a $X_{+90}$ gate creates a superposition state of the control qubit, followed by the $\CR$ gate before the joint readout is used to perform state tomography and reconstruct the two-qubit density matrix $\rho$. The joint readout technique has been shown to be capable of measuring ensembles of both separable and highly-entangled two-qubit states~\cite{Chow2010}. The joint readout assumes the measurement ensemble to be $\langle M \rangle = \beta_{II}+\beta_{IZ} \langle IZ \rangle +\beta_{ZI} \langle ZI \rangle + \beta_{ZZ} \langle ZZ \rangle$. Calibration of the readout gives $[\beta_{II},\beta_{IZ}, \beta_{ZI}, \beta_{ZZ}] = [1,0.77, 0.72,0.6]$. We use maximum-likelihood estimation to extract $\rho$ from a set of experiments involving 15 different single-qubit operations applied to a two-qubit state right before the measurement. 

A standard metric of entanglement, the concurrence $\mathcal{C}$, can be computed for measured $\rho$ generated with our gate protocol for different $\tg$ and $A$. Figures~\ref{fig:3}(b--e) show the evolution of $\mathcal{C}$ with $\tg$ for four different $A$. We find that $\mathcal{C}$ oscillates with a period of $1/\Jeff$. The points of maximal $\mathcal{C}$ correspond to $\tg = 1/2\Jeff$, where the $\CR$ is a $[ZX]_{+90}$ two-qubit operation which produces maximally entangled states in the Bell basis. The solid lines in Fig.~\ref{fig:3}(b--e) correspond to master-equation two-level simulations taking into account the gate  and  coherence times. 

\begin{figure}[t!]
\centering
\includegraphics[width=0.45\textwidth]{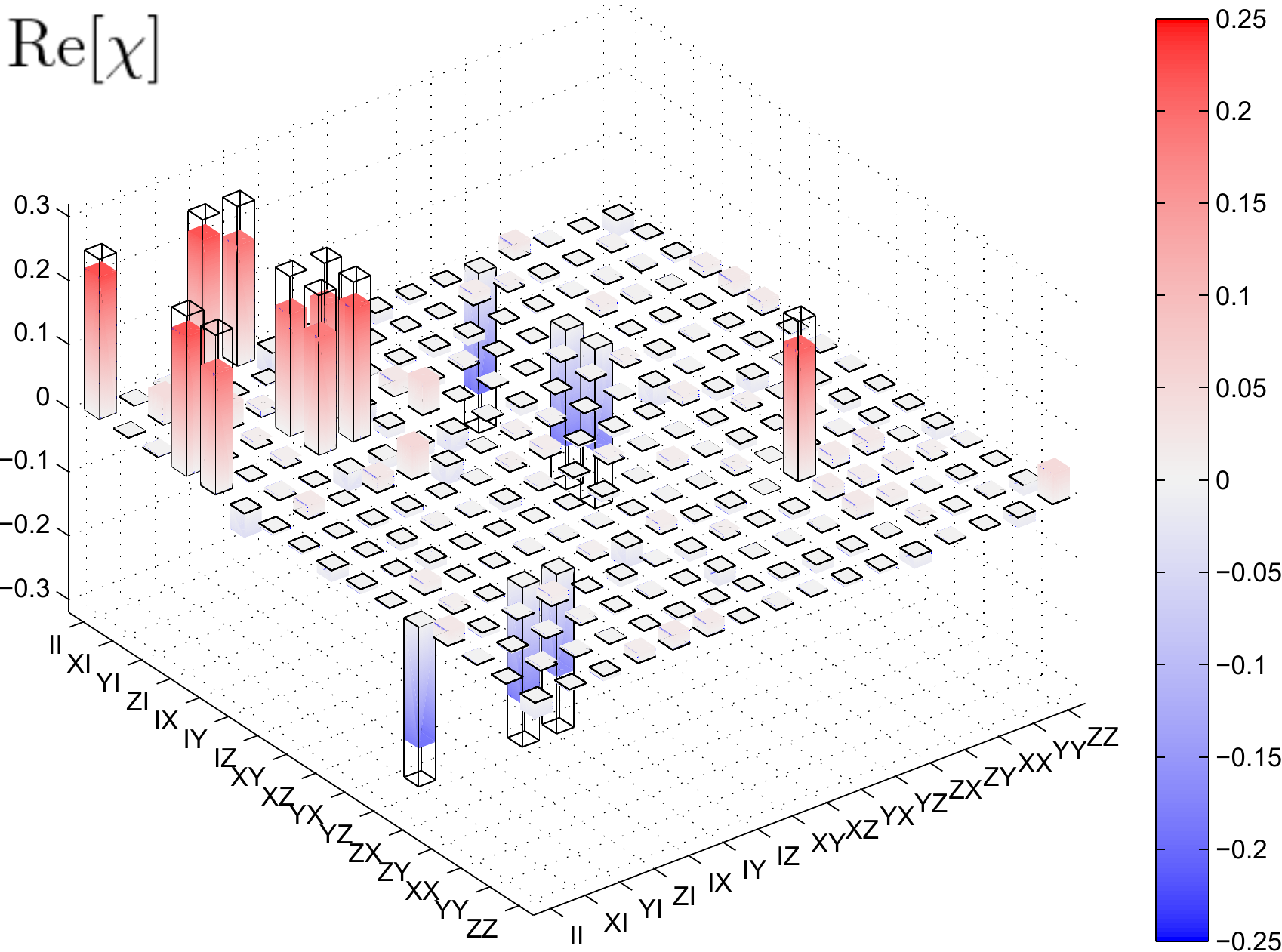}
\caption{(color online)~Quantum process tomography. $\text{Re}[\chi]$ for the optimal $\CR$ gate are shown as the shaded and colored bars, corresponding to $t_{\text{gate}} = 220\,\ns$ and $A'$. The $x$- and $y$-axes are labeled in the two-qubit Pauli operator basis $\{I, X,Y,Z\}^{\otimes 2}$. The ideal two-qubit gate corresponds to a CNOT two-qubit unitary, and the corresponding $\text{Re}[\chi_{\text{ideal}}]$ and $\text{Im}[\chi_{\text{ideal}}]$ are shown as the transparent bars.  All $\text{Im}[\chi]$ bars (not shown) are $ < 0.05$.}
\label{fig:4}
\end{figure}

In Fig.~\ref{fig:3}(f) we show a measured $\rho$ for one of the maximally entangled Bell basis states $\ket{\psi_{\text{Bell}}} = 1/\sqrt{2} (\ket{00} +\ket{11})$, generated with a $\CR$ gate at $\tg= 220\,\ns$ and the amplitude $A/2\pi=553\,\MHz$ which gives the maximal $\mathcal{C}$ in the oscillations shown in Fig.~\ref{fig:3}(e). As previously mentioned, due to the spurious crosstalk on qubit 2 during the gate, an additional single qubit rotation of qubit 2 is usually performed. Although this extra rotation can be simply un-done with an additional single-qubit gate, for this specific $\tg$ and $A$, the additional rotation from the crosstalk is $X_{+90}$, which when combined with the $[ZX]_{+90}$ leaves the two qubits in the canonical Bell state $\ket{\psi_{\text{Bell}}}$. The fidelity of this measured state to the ideal $\ket{\psi_{\text{Bell}}}$ is found to be $\mathcal{F} = \langle \psi_{\text{Bell}}|\rho|\psi_{\text{Bell}}\rangle = 90\%\pm0.04$ with a concurrence of $\mathcal{C} = 0.88\pm 0.05$.

The $\CR$ gate is finally characterized using quantum process tomography~\cite{nielsen_chuang_2000} (QPT). First, we create the input states corresponding to applying combinations of single-qubit gates $\{I, X_{\pm90}, Y_{\pm90}, X\}$ on both qubits. Then we operate $\CR(A,\tg)$ on all 36 such input states and perform state tomography. The process matrix $\chi$ is obtained and compared to the ideal $\chi_{\text{ideal}}$ (see Fig.~\ref{fig:4}) to give a process fidelity $\mathcal{F}_p = 0.77$ and a gate fidelity~\cite{nielsen_gatefid_2002} $\mathcal{F}_{\text{g}} = 0.81$, which is consistent with a simulated gate fidelity of 0.86 that takes into account the measured coherence times. The difference in the values is attributable to calibration errors on the single-qubit preparation and analysis gates. As an experimental measure of the effectiveness of the $\CR$ gate we also perform QPT for a 220~ns identity operation, where we find $\mathcal{F}_{\text{g}} = 0.81$, which critically is the same as the $\CR$ gate fidelity. For a test of other residual two-qubit interactions in the system, we extract a maximum $\mathcal{C}=0.09$ from the action of the identity operation over all separable input states. This is consistent with a measured residual $ZZ$ interaction of 200 kHz, which is an effect common to the circuit QED system~\cite{dicarlo_2009}.

Thus, we have developed a microwaves-only scheme for a two-qubit universal gate capable of generating highly entangled states with superconducting qubits. Furthermore, the underlying two-qubit interaction is tunable simply via increasing the amplitude of a microwave drive. Although we saturate to a maximal interaction strength in this work [Fig.~\ref{fig:2}(b)], we anticipate surpassing this limit with additional pulse shaping on the cross-resonance drive in the future. The cross-resonance coupling protocol is minimal in complexity to implement as it requires no additional subcircuits or controls other than those for addressing each qubit independently. The gate can be immediately expanded to generate maximally entangled states for systems of more than two fixed-frequency qubits and to couple non-nearest neighbor qubits in frequency. The cross-resonance protocol is therefore poised to be a useful experimental tool for larger-scale quantum information processors.

\begin{acknowledgments}
	We thank Kent Fung and Jack Rohrs for experimental contributions and Marcus P.~da Silva, Zachary Dutton, and Graeme Smith for engaging theoretical discussions. We acknowledge support from IARPA under contract W911NF-10-1-0324. All statements of fact, opinion or conclusions contained herein are those of the authors and should not be construed as representing the official views or policies of the U.S. Government.
\end{acknowledgments}


\begin{thebibliography}{27}%
\makeatletter
\providecommand \@ifxundefined [1]{%
 \@ifx{#1\undefined}
}%
\providecommand \@ifnum [1]{%
 \ifnum #1\expandafter \@firstoftwo
 \else \expandafter \@secondoftwo
 \fi
}%
\providecommand \@ifx [1]{%
 \ifx #1\expandafter \@firstoftwo
 \else \expandafter \@secondoftwo
 \fi
}%
\providecommand \natexlab [1]{#1}%
\providecommand \enquote  [1]{``#1''}%
\providecommand \bibnamefont  [1]{#1}%
\providecommand \bibfnamefont [1]{#1}%
\providecommand \citenamefont [1]{#1}%
\providecommand \href@noop [0]{\@secondoftwo}%
\providecommand \href [0]{\begingroup \@sanitize@url \@href}%
\providecommand \@href[1]{\@@startlink{#1}\@@href}%
\providecommand \@@href[1]{\endgroup#1\@@endlink}%
\providecommand \@sanitize@url [0]{\catcode `\\12\catcode `\$12\catcode
  `\&12\catcode `\#12\catcode `\^12\catcode `\_12\catcode `\%12\relax}%
\providecommand \@@startlink[1]{}%
\providecommand \@@endlink[0]{}%
\providecommand \url  [0]{\begingroup\@sanitize@url \@url }%
\providecommand \@url [1]{\endgroup\@href {#1}{\urlprefix }}%
\providecommand \urlprefix  [0]{URL }%
\providecommand \Eprint [0]{\href }%
\providecommand \doibase [0]{http://dx.doi.org/}%
\providecommand \selectlanguage [0]{\@gobble}%
\providecommand \bibinfo  [0]{\@secondoftwo}%
\providecommand \bibfield  [0]{\@secondoftwo}%
\providecommand \translation [1]{[#1]}%
\providecommand \BibitemOpen [0]{}%
\providecommand \bibitemStop [0]{}%
\providecommand \bibitemNoStop [0]{.\EOS\space}%
\providecommand \EOS [0]{\spacefactor3000\relax}%
\providecommand \BibitemShut  [1]{\csname bibitem#1\endcsname}%
\let\auto@bib@innerbib\@empty
%</preamble>
\bibitem [{\citenamefont {Lucero}\ \emph {et~al.}(2008)\citenamefont {Lucero},
  \citenamefont {Hofheinz}, \citenamefont {Ansmann}, \citenamefont {Bialczak},
  \citenamefont {Katz}, \citenamefont {Neeley}, \citenamefont {O'Connell},
  \citenamefont {Wang}, \citenamefont {Cleland},\ and\ \citenamefont
  {Martinis}}]{lucero_gates_2008}%
  \BibitemOpen
  \bibfield  {author} {\bibinfo {author} {\bibfnamefont {E.}~\bibnamefont
  {Lucero}}, \bibinfo {author} {\bibfnamefont {M.}~\bibnamefont {Hofheinz}},
  \bibinfo {author} {\bibfnamefont {M.}~\bibnamefont {Ansmann}}, \bibinfo
  {author} {\bibfnamefont {R.~C.}\ \bibnamefont {Bialczak}}, \bibinfo {author}
  {\bibfnamefont {N.}~\bibnamefont {Katz}}, \bibinfo {author} {\bibfnamefont
  {M.}~\bibnamefont {Neeley}}, \bibinfo {author} {\bibfnamefont {A.~D.}\
  \bibnamefont {O'Connell}}, \bibinfo {author} {\bibfnamefont {H.}~\bibnamefont
  {Wang}}, \bibinfo {author} {\bibfnamefont {A.~N.}\ \bibnamefont {Cleland}}, \
  and\ \bibinfo {author} {\bibfnamefont {J.~M.}\ \bibnamefont {Martinis}},\
  }\href {\doibase 10.1103/PhysRevLett.100.247001} {\bibfield  {journal}
  {\bibinfo  {journal} {Phys. Rev. Lett.}\ }\textbf {\bibinfo {volume} {100}},\
  \bibinfo {eid} {247001} (\bibinfo {year} {2008})}\BibitemShut {NoStop}%
\bibitem [{\citenamefont {Chow}\ \emph
  {et~al.}(2010{\natexlab{a}})\citenamefont {Chow}, \citenamefont {DiCarlo},
  \citenamefont {Gambetta}, \citenamefont {Motzoi}, \citenamefont {Frunzio},
  \citenamefont {Girvin},\ and\ \citenamefont {Schoelkopf}}]{Chow2010a}%
  \BibitemOpen
  \bibfield  {author} {\bibinfo {author} {\bibfnamefont {J.~M.}\ \bibnamefont
  {Chow}}, \bibinfo {author} {\bibfnamefont {L.}~\bibnamefont {DiCarlo}},
  \bibinfo {author} {\bibfnamefont {J.~M.}\ \bibnamefont {Gambetta}}, \bibinfo
  {author} {\bibfnamefont {F.}~\bibnamefont {Motzoi}}, \bibinfo {author}
  {\bibfnamefont {L.}~\bibnamefont {Frunzio}}, \bibinfo {author} {\bibfnamefont
  {S.~M.}\ \bibnamefont {Girvin}}, \ and\ \bibinfo {author} {\bibfnamefont
  {R.~J.}\ \bibnamefont {Schoelkopf}},\ }\href
  {http://link.aps.org/doi/10.1103/PhysRevA.82.040305} {\bibfield  {journal}
  {\bibinfo  {journal} {Phys. Rev. A}\ }\textbf {\bibinfo {volume} {82}},\
  \bibinfo {pages} {040305} (\bibinfo {year} {2010}{\natexlab{a}})}\BibitemShut
  {NoStop}%
\bibitem [{\citenamefont {Plantenberg}\ \emph {et~al.}(2007)\citenamefont
  {Plantenberg}, \citenamefont {de~Groot}, \citenamefont {Harmans},\ and\
  \citenamefont {Mooij}}]{Plantenberg:2007it}%
  \BibitemOpen
  \bibfield  {author} {\bibinfo {author} {\bibfnamefont {J.~H.}\ \bibnamefont
  {Plantenberg}}, \bibinfo {author} {\bibfnamefont {P.~C.}\ \bibnamefont
  {de~Groot}}, \bibinfo {author} {\bibfnamefont {C.~J. P.~M.}\ \bibnamefont
  {Harmans}}, \ and\ \bibinfo {author} {\bibfnamefont {J.~E.}\ \bibnamefont
  {Mooij}},\ }\href {\doibase 10.1038/nature05896} {\bibfield  {journal}
  {\bibinfo  {journal} {Nature}\ }\textbf {\bibinfo {volume} {447}},\ \bibinfo
  {pages} {836} (\bibinfo {year} {2007})}\BibitemShut {NoStop}%
\bibitem [{\citenamefont {DiCarlo}\ \emph {et~al.}(2009)\citenamefont
  {DiCarlo}, \citenamefont {Chow}, \citenamefont {Gambetta}, \citenamefont
  {Bishop}, \citenamefont {Johnson}, \citenamefont {Schuster}, \citenamefont
  {Majer}, \citenamefont {Blais}, \citenamefont {Frunzio}, \citenamefont
  {Girvin},\ and\ \citenamefont {Schoelkopf}}]{dicarlo_2009}%
  \BibitemOpen
  \bibfield  {author} {\bibinfo {author} {\bibfnamefont {L.}~\bibnamefont
  {DiCarlo}}, \bibinfo {author} {\bibfnamefont {J.~M.}\ \bibnamefont {Chow}},
  \bibinfo {author} {\bibfnamefont {J.~M.}\ \bibnamefont {Gambetta}}, \bibinfo
  {author} {\bibfnamefont {L.~S.}\ \bibnamefont {Bishop}}, \bibinfo {author}
  {\bibfnamefont {B.~R.}\ \bibnamefont {Johnson}}, \bibinfo {author}
  {\bibfnamefont {D.~I.}\ \bibnamefont {Schuster}}, \bibinfo {author}
  {\bibfnamefont {J.}~\bibnamefont {Majer}}, \bibinfo {author} {\bibfnamefont
  {A.}~\bibnamefont {Blais}}, \bibinfo {author} {\bibfnamefont
  {L.}~\bibnamefont {Frunzio}}, \bibinfo {author} {\bibfnamefont {S.~M.}\
  \bibnamefont {Girvin}}, \ and\ \bibinfo {author} {\bibfnamefont {R.~J.}\
  \bibnamefont {Schoelkopf}},\ }\href {\doibase 10.1038/nature08121} {\bibfield
   {journal} {\bibinfo  {journal} {Nature}\ }\textbf {\bibinfo {volume}
  {460}},\ \bibinfo {pages} {240} (\bibinfo {year} {2009})}\BibitemShut
  {NoStop}%
\bibitem [{\citenamefont {Ansmann}\ \emph {et~al.}(2009)\citenamefont
  {Ansmann}, \citenamefont {Wang}, \citenamefont {Bialczak}, \citenamefont
  {Hofheinz}, \citenamefont {Lucero}, \citenamefont {Neeley}, \citenamefont
  {O'Connell}, \citenamefont {Sank}, \citenamefont {Weides}, \citenamefont
  {Wenner}, \citenamefont {Cleland},\ and\ \citenamefont
  {Martinis}}]{Ansmann:2009yq}%
  \BibitemOpen
  \bibfield  {author} {\bibinfo {author} {\bibfnamefont {M.}~\bibnamefont
  {Ansmann}}, \bibinfo {author} {\bibfnamefont {H.}~\bibnamefont {Wang}},
  \bibinfo {author} {\bibfnamefont {R.~C.}\ \bibnamefont {Bialczak}}, \bibinfo
  {author} {\bibfnamefont {M.}~\bibnamefont {Hofheinz}}, \bibinfo {author}
  {\bibfnamefont {E.}~\bibnamefont {Lucero}}, \bibinfo {author} {\bibfnamefont
  {M.}~\bibnamefont {Neeley}}, \bibinfo {author} {\bibfnamefont {A.~D.}\
  \bibnamefont {O'Connell}}, \bibinfo {author} {\bibfnamefont {D.}~\bibnamefont
  {Sank}}, \bibinfo {author} {\bibfnamefont {M.}~\bibnamefont {Weides}},
  \bibinfo {author} {\bibfnamefont {J.}~\bibnamefont {Wenner}}, \bibinfo
  {author} {\bibfnamefont {A.~N.}\ \bibnamefont {Cleland}}, \ and\ \bibinfo
  {author} {\bibfnamefont {J.~M.}\ \bibnamefont {Martinis}},\ }\href {\doibase
  10.1038/nature08363} {\bibfield  {journal} {\bibinfo  {journal} {Nature}\
  }\textbf {\bibinfo {volume} {461}},\ \bibinfo {pages} {504} (\bibinfo {year}
  {2009})}\BibitemShut {NoStop}%
\bibitem [{\citenamefont {DiCarlo}\ \emph {et~al.}(2010)\citenamefont
  {DiCarlo}, \citenamefont {Reed}, \citenamefont {Sun}, \citenamefont
  {Johnson}, \citenamefont {Chow}, \citenamefont {Gambetta}, \citenamefont
  {Frunzio}, \citenamefont {Girvin}, \citenamefont {Devoret},\ and\
  \citenamefont {Schoelkopf}}]{DiCarlo2010}%
  \BibitemOpen
  \bibfield  {author} {\bibinfo {author} {\bibfnamefont {L.}~\bibnamefont
  {DiCarlo}}, \bibinfo {author} {\bibfnamefont {M.~D.}\ \bibnamefont {Reed}},
  \bibinfo {author} {\bibfnamefont {L.}~\bibnamefont {Sun}}, \bibinfo {author}
  {\bibfnamefont {B.~R.}\ \bibnamefont {Johnson}}, \bibinfo {author}
  {\bibfnamefont {J.~M.}\ \bibnamefont {Chow}}, \bibinfo {author}
  {\bibfnamefont {J.~M.}\ \bibnamefont {Gambetta}}, \bibinfo {author}
  {\bibfnamefont {L.}~\bibnamefont {Frunzio}}, \bibinfo {author} {\bibfnamefont
  {S.~M.}\ \bibnamefont {Girvin}}, \bibinfo {author} {\bibfnamefont {M.~H.}\
  \bibnamefont {Devoret}}, \ and\ \bibinfo {author} {\bibfnamefont {R.~J.}\
  \bibnamefont {Schoelkopf}},\ }\href {http://dx.doi.org/10.1038/nature09416}
  {\bibfield  {journal} {\bibinfo  {journal} {Nature}\ }\textbf {\bibinfo
  {volume} {467}},\ \bibinfo {pages} {574} (\bibinfo {year}
  {2010})}\BibitemShut {NoStop}%
\bibitem [{\citenamefont {Neeley}\ \emph {et~al.}(2010)\citenamefont {Neeley},
  \citenamefont {Bialczak}, \citenamefont {Lenander}, \citenamefont {Lucero},
  \citenamefont {Mariantoni}, \citenamefont {O/'Connell}, \citenamefont {Sank},
  \citenamefont {Wang}, \citenamefont {Weides}, \citenamefont {Wenner},
  \citenamefont {Yin}, \citenamefont {Yamamoto}, \citenamefont {Cleland},\ and\
  \citenamefont {Martinis}}]{Neeley2010}%
  \BibitemOpen
  \bibfield  {author} {\bibinfo {author} {\bibfnamefont {M.}~\bibnamefont
  {Neeley}}, \bibinfo {author} {\bibfnamefont {R.~C.}\ \bibnamefont
  {Bialczak}}, \bibinfo {author} {\bibfnamefont {M.}~\bibnamefont {Lenander}},
  \bibinfo {author} {\bibfnamefont {E.}~\bibnamefont {Lucero}}, \bibinfo
  {author} {\bibfnamefont {M.}~\bibnamefont {Mariantoni}}, \bibinfo {author}
  {\bibfnamefont {A.~D.}\ \bibnamefont {O'Connell}}, \bibinfo {author}
  {\bibfnamefont {D.}~\bibnamefont {Sank}}, \bibinfo {author} {\bibfnamefont
  {H.}~\bibnamefont {Wang}}, \bibinfo {author} {\bibfnamefont {M.}~\bibnamefont
  {Weides}}, \bibinfo {author} {\bibfnamefont {J.}~\bibnamefont {Wenner}},
  \bibinfo {author} {\bibfnamefont {Y.}~\bibnamefont {Yin}}, \bibinfo {author}
  {\bibfnamefont {T.}~\bibnamefont {Yamamoto}}, \bibinfo {author}
  {\bibfnamefont {A.~N.}\ \bibnamefont {Cleland}}, \ and\ \bibinfo {author}
  {\bibfnamefont {J.~M.}\ \bibnamefont {Martinis}},\ }\href
  {http://dx.doi.org/10.1038/nature09418} {\bibfield  {journal} {\bibinfo
  {journal} {Nature}\ }\textbf {\bibinfo {volume} {467}},\ \bibinfo {pages}
  {570} (\bibinfo {year} {2010})}\BibitemShut {NoStop}%
\bibitem [{\citenamefont {Bertet}\ \emph {et~al.}(2006)\citenamefont {Bertet},
  \citenamefont {Harmans},\ and\ \citenamefont {Mooij}}]{Bertet2006}%
  \BibitemOpen
  \bibfield  {author} {\bibinfo {author} {\bibfnamefont {P.}~\bibnamefont
  {Bertet}}, \bibinfo {author} {\bibfnamefont {C.~J. P.~M.}\ \bibnamefont
  {Harmans}}, \ and\ \bibinfo {author} {\bibfnamefont {J.~E.}\ \bibnamefont
  {Mooij}},\ }\href {http://link.aps.org/doi/10.1103/PhysRevB.73.064512}
  {\bibfield  {journal} {\bibinfo  {journal} {Phys. Rev. B}\ }\textbf {\bibinfo
  {volume} {73}},\ \bibinfo {pages} {064512} (\bibinfo {year}
  {2006})}\BibitemShut {NoStop}%
\bibitem [{\citenamefont {Niskanen}\ \emph {et~al.}(2007)\citenamefont
  {Niskanen}, \citenamefont {Harrabi}, \citenamefont {Yoshihara}, \citenamefont
  {Nakamura}, \citenamefont {Lloyd},\ and\ \citenamefont
  {Tsai}}]{Niskanen:2007if}%
  \BibitemOpen
  \bibfield  {author} {\bibinfo {author} {\bibfnamefont {A.~O.}\ \bibnamefont
  {Niskanen}}, \bibinfo {author} {\bibfnamefont {K.}~\bibnamefont {Harrabi}},
  \bibinfo {author} {\bibfnamefont {F.}~\bibnamefont {Yoshihara}}, \bibinfo
  {author} {\bibfnamefont {Y.}~\bibnamefont {Nakamura}}, \bibinfo {author}
  {\bibfnamefont {S.}~\bibnamefont {Lloyd}}, \ and\ \bibinfo {author}
  {\bibfnamefont {J.~S.}\ \bibnamefont {Tsai}},\ }\href {\doibase
  10.1126/science.1141324} {\bibfield  {journal} {\bibinfo  {journal}
  {Science}\ }\textbf {\bibinfo {volume} {316}},\ \bibinfo {pages} {723}
  (\bibinfo {year} {2007})}\BibitemShut {NoStop}%
\bibitem [{\citenamefont {Harris}\ \emph {et~al.}(2007)\citenamefont {Harris},
  \citenamefont {Berkley}, \citenamefont {Johnson}, \citenamefont {Bunyk},
  \citenamefont {Govorkov}, \citenamefont {Thom}, \citenamefont {Uchaikin},
  \citenamefont {Wilson}, \citenamefont {Chung}, \citenamefont {Holtham},
  \citenamefont {Biamonte}, \citenamefont {Smirnov}, \citenamefont {Amin},\
  and\ \citenamefont {Maassen van~den Brink}}]{Harris2007}%
  \BibitemOpen
  \bibfield  {author} {\bibinfo {author} {\bibfnamefont {R.}~\bibnamefont
  {Harris}}, \bibinfo {author} {\bibfnamefont {A.~J.}\ \bibnamefont {Berkley}},
  \bibinfo {author} {\bibfnamefont {M.~W.}\ \bibnamefont {Johnson}}, \bibinfo
  {author} {\bibfnamefont {P.}~\bibnamefont {Bunyk}}, \bibinfo {author}
  {\bibfnamefont {S.}~\bibnamefont {Govorkov}}, \bibinfo {author}
  {\bibfnamefont {M.~C.}\ \bibnamefont {Thom}}, \bibinfo {author}
  {\bibfnamefont {S.}~\bibnamefont {Uchaikin}}, \bibinfo {author}
  {\bibfnamefont {A.~B.}\ \bibnamefont {Wilson}}, \bibinfo {author}
  {\bibfnamefont {J.}~\bibnamefont {Chung}}, \bibinfo {author} {\bibfnamefont
  {E.}~\bibnamefont {Holtham}}, \bibinfo {author} {\bibfnamefont {J.~D.}\
  \bibnamefont {Biamonte}}, \bibinfo {author} {\bibfnamefont {A.~Y.}\
  \bibnamefont {Smirnov}}, \bibinfo {author} {\bibfnamefont {M.~H.~S.}\
  \bibnamefont {Amin}}, \ and\ \bibinfo {author} {\bibfnamefont
  {A.}~\bibnamefont {Maassen van~den Brink}},\ }\href
  {http://link.aps.org/doi/10.1103/PhysRevLett.98.177001} {\bibfield  {journal}
  {\bibinfo  {journal} {Phys. Rev. Lett.}\ }\textbf {\bibinfo {volume} {98}},\
  \bibinfo {pages} {177001} (\bibinfo {year} {2007})}\BibitemShut {NoStop}%
\bibitem [{\citenamefont {Hime}\ \emph {et~al.}(2006)\citenamefont {Hime},
  \citenamefont {Reichardt}, \citenamefont {Plourde}, \citenamefont
  {Robertson}, \citenamefont {Wu}, \citenamefont {Ustinov},\ and\ \citenamefont
  {Clarke}}]{Hime:2006ef}%
  \BibitemOpen
  \bibfield  {author} {\bibinfo {author} {\bibfnamefont {T.}~\bibnamefont
  {Hime}}, \bibinfo {author} {\bibfnamefont {P.~A.}\ \bibnamefont {Reichardt}},
  \bibinfo {author} {\bibfnamefont {B.~L.~T.}\ \bibnamefont {Plourde}},
  \bibinfo {author} {\bibfnamefont {T.~L.}\ \bibnamefont {Robertson}}, \bibinfo
  {author} {\bibfnamefont {C.~E.}\ \bibnamefont {Wu}}, \bibinfo {author}
  {\bibfnamefont {A.~V.}\ \bibnamefont {Ustinov}}, \ and\ \bibinfo {author}
  {\bibfnamefont {J.}~\bibnamefont {Clarke}},\ }\href {\doibase
  10.1126/science.1134388} {\bibfield  {journal} {\bibinfo  {journal}
  {Science}\ }\textbf {\bibinfo {volume} {314}},\ \bibinfo {pages} {1427}
  (\bibinfo {year} {2006})}\BibitemShut {NoStop}%
\bibitem [{\citenamefont {Bialczak}\ \emph {et~al.}(2011)\citenamefont
  {Bialczak}, \citenamefont {Ansmann}, \citenamefont {Hofheinz}, \citenamefont
  {Lenander}, \citenamefont {Lucero}, \citenamefont {Neeley}, \citenamefont
  {O'Connell}, \citenamefont {Sank}, \citenamefont {Wang}, \citenamefont
  {Weides}, \citenamefont {Wenner}, \citenamefont {Yamamoto}, \citenamefont
  {Cleland},\ and\ \citenamefont {Martinis}}]{Bialczak2011}%
  \BibitemOpen
  \bibfield  {author} {\bibinfo {author} {\bibfnamefont {R.~C.}\ \bibnamefont
  {Bialczak}}, \bibinfo {author} {\bibfnamefont {M.}~\bibnamefont {Ansmann}},
  \bibinfo {author} {\bibfnamefont {M.}~\bibnamefont {Hofheinz}}, \bibinfo
  {author} {\bibfnamefont {M.}~\bibnamefont {Lenander}}, \bibinfo {author}
  {\bibfnamefont {E.}~\bibnamefont {Lucero}}, \bibinfo {author} {\bibfnamefont
  {M.}~\bibnamefont {Neeley}}, \bibinfo {author} {\bibfnamefont {A.~D.}\
  \bibnamefont {O'Connell}}, \bibinfo {author} {\bibfnamefont {D.}~\bibnamefont
  {Sank}}, \bibinfo {author} {\bibfnamefont {H.}~\bibnamefont {Wang}}, \bibinfo
  {author} {\bibfnamefont {M.}~\bibnamefont {Weides}}, \bibinfo {author}
  {\bibfnamefont {J.}~\bibnamefont {Wenner}}, \bibinfo {author} {\bibfnamefont
  {T.}~\bibnamefont {Yamamoto}}, \bibinfo {author} {\bibfnamefont {A.~N.}\
  \bibnamefont {Cleland}}, \ and\ \bibinfo {author} {\bibfnamefont {J.~M.}\
  \bibnamefont {Martinis}},\ }\href
  {http://link.aps.org/doi/10.1103/PhysRevLett.106.060501} {\bibfield
  {journal} {\bibinfo  {journal} {Phys. Rev. Lett.}\ }\textbf {\bibinfo
  {volume} {106}},\ \bibinfo {pages} {060501} (\bibinfo {year}
  {2011})}\BibitemShut {NoStop}%
\bibitem [{\citenamefont {Srinivasan}\ \emph {et~al.}(2011)\citenamefont
  {Srinivasan}, \citenamefont {Hoffman}, \citenamefont {Gambetta},\ and\
  \citenamefont {Houck}}]{Sri2011}%
  \BibitemOpen
  \bibfield  {author} {\bibinfo {author} {\bibfnamefont {S.~J.}\ \bibnamefont
  {Srinivasan}}, \bibinfo {author} {\bibfnamefont {A.~J.}\ \bibnamefont
  {Hoffman}}, \bibinfo {author} {\bibfnamefont {J.~M.}\ \bibnamefont
  {Gambetta}}, \ and\ \bibinfo {author} {\bibfnamefont {A.~A.}\ \bibnamefont
  {Houck}},\ }\href {http://link.aps.org/doi/10.1103/PhysRevLett.106.083601}
  {\bibfield  {journal} {\bibinfo  {journal} {Phys. Rev. Lett.}\ }\textbf
  {\bibinfo {volume} {106}},\ \bibinfo {pages} {083601} (\bibinfo {year}
  {2011})}\BibitemShut {NoStop}%
\bibitem [{\citenamefont {Steffen}\ \emph {et~al.}(2006)\citenamefont
  {Steffen}, \citenamefont {Ansmann}, \citenamefont {Bialczak}, \citenamefont
  {Katz}, \citenamefont {Lucero}, \citenamefont {McDermott}, \citenamefont
  {Neeley}, \citenamefont {Weig}, \citenamefont {Cleland},\ and\ \citenamefont
  {Martinis}}]{steffen_entang_2006}%
  \BibitemOpen
  \bibfield  {author} {\bibinfo {author} {\bibfnamefont {M.}~\bibnamefont
  {Steffen}}, \bibinfo {author} {\bibfnamefont {M.}~\bibnamefont {Ansmann}},
  \bibinfo {author} {\bibfnamefont {R.~C.}\ \bibnamefont {Bialczak}}, \bibinfo
  {author} {\bibfnamefont {N.}~\bibnamefont {Katz}}, \bibinfo {author}
  {\bibfnamefont {E.}~\bibnamefont {Lucero}}, \bibinfo {author} {\bibfnamefont
  {R.}~\bibnamefont {McDermott}}, \bibinfo {author} {\bibfnamefont
  {M.}~\bibnamefont {Neeley}}, \bibinfo {author} {\bibfnamefont {E.~M.}\
  \bibnamefont {Weig}}, \bibinfo {author} {\bibfnamefont {A.~N.}\ \bibnamefont
  {Cleland}}, \ and\ \bibinfo {author} {\bibfnamefont {J.~M.}\ \bibnamefont
  {Martinis}},\ }\href {\doibase 10.1126/science.1130886} {\bibfield  {journal}
  {\bibinfo  {journal} {Science}\ }\textbf {\bibinfo {volume} {313}},\ \bibinfo
  {pages} {1423} (\bibinfo {year} {2006})}\BibitemShut {NoStop}%
\bibitem [{\citenamefont {Bialczak}\ \emph {et~al.}(2010)\citenamefont
  {Bialczak}, \citenamefont {Ansmann}, \citenamefont {Hofheinz}, \citenamefont
  {Lucero}, \citenamefont {Neeley}, \citenamefont {O/'Connell}, \citenamefont
  {Sank}, \citenamefont {Wang}, \citenamefont {Wenner}, \citenamefont
  {Steffen}, \citenamefont {Cleland},\ and\ \citenamefont
  {Martinis}}]{Bialczak2010}%
  \BibitemOpen
  \bibfield  {author} {\bibinfo {author} {\bibfnamefont {R.~C.}\ \bibnamefont
  {Bialczak}}, \bibinfo {author} {\bibfnamefont {M.}~\bibnamefont {Ansmann}},
  \bibinfo {author} {\bibfnamefont {M.}~\bibnamefont {Hofheinz}}, \bibinfo
  {author} {\bibfnamefont {E.}~\bibnamefont {Lucero}}, \bibinfo {author}
  {\bibfnamefont {M.}~\bibnamefont {Neeley}}, \bibinfo {author} {\bibfnamefont
  {A.~D.}\ \bibnamefont {O/'Connell}}, \bibinfo {author} {\bibfnamefont
  {D.}~\bibnamefont {Sank}}, \bibinfo {author} {\bibfnamefont {H.}~\bibnamefont
  {Wang}}, \bibinfo {author} {\bibfnamefont {J.}~\bibnamefont {Wenner}},
  \bibinfo {author} {\bibfnamefont {M.}~\bibnamefont {Steffen}}, \bibinfo
  {author} {\bibfnamefont {A.~N.}\ \bibnamefont {Cleland}}, \ and\ \bibinfo
  {author} {\bibfnamefont {J.~M.}\ \bibnamefont {Martinis}},\ }\href
  {http://dx.doi.org/10.1038/nphys1639} {\bibfield  {journal} {\bibinfo
  {journal} {Nat Phys}\ }\textbf {\bibinfo {volume} {6}},\ \bibinfo {pages}
  {409} (\bibinfo {year} {2010})}\BibitemShut {NoStop}%
\bibitem [{\citenamefont {Paraoanu}(2006)}]{Paraoanu2006}%
  \BibitemOpen
  \bibfield  {author} {\bibinfo {author} {\bibfnamefont {G.~S.}\ \bibnamefont
  {Paraoanu}},\ }\href {http://link.aps.org/doi/10.1103/PhysRevB.74.140504}
  {\bibfield  {journal} {\bibinfo  {journal} {Phys. Rev. B}\ }\textbf {\bibinfo
  {volume} {74}},\ \bibinfo {pages} {140504} (\bibinfo {year}
  {2006})}\BibitemShut {NoStop}%
\bibitem [{\citenamefont {Rigetti}\ and\ \citenamefont
  {Devoret}(2010)}]{Rigetti2010}%
  \BibitemOpen
  \bibfield  {author} {\bibinfo {author} {\bibfnamefont {C.}~\bibnamefont
  {Rigetti}}\ and\ \bibinfo {author} {\bibfnamefont {M.}~\bibnamefont
  {Devoret}},\ }\href {http://link.aps.org/doi/10.1103/PhysRevB.81.134507}
  {\bibfield  {journal} {\bibinfo  {journal} {Phys. Rev. B}\ }\textbf {\bibinfo
  {volume} {81}},\ \bibinfo {pages} {134507} (\bibinfo {year}
  {2010})}\BibitemShut {NoStop}%
\bibitem [{\citenamefont {de~Groot}\ \emph {et~al.}(2010)\citenamefont
  {de~Groot}, \citenamefont {Lisenfeld}, \citenamefont {Schouten},
  \citenamefont {Ashhab}, \citenamefont {Lupascu}, \citenamefont {Harmans},\
  and\ \citenamefont {Mooij}}]{Groot2010}%
  \BibitemOpen
  \bibfield  {author} {\bibinfo {author} {\bibfnamefont {P.~C.}\ \bibnamefont
  {de~Groot}}, \bibinfo {author} {\bibfnamefont {J.}~\bibnamefont {Lisenfeld}},
  \bibinfo {author} {\bibfnamefont {R.~N.}\ \bibnamefont {Schouten}}, \bibinfo
  {author} {\bibfnamefont {S.}~\bibnamefont {Ashhab}}, \bibinfo {author}
  {\bibfnamefont {A.}~\bibnamefont {Lupascu}}, \bibinfo {author} {\bibfnamefont
  {C.~J. P.~M.}\ \bibnamefont {Harmans}}, \ and\ \bibinfo {author}
  {\bibfnamefont {J.~E.}\ \bibnamefont {Mooij}},\ }\href
  {http://dx.doi.org/10.1038/nphys1733} {\bibfield  {journal} {\bibinfo
  {journal} {Nat Phys}\ }\textbf {\bibinfo {volume} {6}},\ \bibinfo {pages}
  {763} (\bibinfo {year} {2010})}\BibitemShut {NoStop}%
\bibitem [{\citenamefont {Majer}\ \emph {et~al.}(2007)\citenamefont {Majer},
  \citenamefont {Chow}, \citenamefont {Gambetta}, \citenamefont {Koch},
  \citenamefont {Johnson}, \citenamefont {Schreier}, \citenamefont {Frunzio},
  \citenamefont {Schuster}, \citenamefont {Houck}, \citenamefont {Wallraff},
  \citenamefont {Blais}, \citenamefont {Devoret}, \citenamefont {Girvin},\ and\
  \citenamefont {Schoelkopf}}]{majer_coupling_2007}%
  \BibitemOpen
  \bibfield  {author} {\bibinfo {author} {\bibfnamefont {J.}~\bibnamefont
  {Majer}}, \bibinfo {author} {\bibfnamefont {J.~M.}\ \bibnamefont {Chow}},
  \bibinfo {author} {\bibfnamefont {J.~M.}\ \bibnamefont {Gambetta}}, \bibinfo
  {author} {\bibfnamefont {J.}~\bibnamefont {Koch}}, \bibinfo {author}
  {\bibfnamefont {B.~R.}\ \bibnamefont {Johnson}}, \bibinfo {author}
  {\bibfnamefont {J.~A.}\ \bibnamefont {Schreier}}, \bibinfo {author}
  {\bibfnamefont {L.}~\bibnamefont {Frunzio}}, \bibinfo {author} {\bibfnamefont
  {D.~I.}\ \bibnamefont {Schuster}}, \bibinfo {author} {\bibfnamefont {A.~A.}\
  \bibnamefont {Houck}}, \bibinfo {author} {\bibfnamefont {A.}~\bibnamefont
  {Wallraff}}, \bibinfo {author} {\bibfnamefont {A.}~\bibnamefont {Blais}},
  \bibinfo {author} {\bibfnamefont {M.~H.}\ \bibnamefont {Devoret}}, \bibinfo
  {author} {\bibfnamefont {S.~M.}\ \bibnamefont {Girvin}}, \ and\ \bibinfo
  {author} {\bibfnamefont {R.~J.}\ \bibnamefont {Schoelkopf}},\ }\href
  {\doibase 10.1038/nature06184} {\bibfield  {journal} {\bibinfo  {journal}
  {Nature}\ }\textbf {\bibinfo {volume} {449}},\ \bibinfo {pages} {443}
  (\bibinfo {year} {2007})}\BibitemShut {NoStop}%
\bibitem [{\citenamefont {Sillanp{\"a}{\"a}}\ \emph {et~al.}(2007)\citenamefont
  {Sillanp{\"a}{\"a}}, \citenamefont {Park},\ and\ \citenamefont
  {Simmonds}}]{Sillanpaa_2007}%
  \BibitemOpen
  \bibfield  {author} {\bibinfo {author} {\bibfnamefont {M.~A.}\ \bibnamefont
  {Sillanp{\"a}{\"a}}}, \bibinfo {author} {\bibfnamefont {J.~I.}\ \bibnamefont
  {Park}}, \ and\ \bibinfo {author} {\bibfnamefont {R.~W.}\ \bibnamefont
  {Simmonds}},\ }\href {\doibase 10.1038/nature06124} {\bibfield  {journal}
  {\bibinfo  {journal} {Nature}\ }\textbf {\bibinfo {volume} {449}},\ \bibinfo
  {pages} {438} (\bibinfo {year} {2007})}\BibitemShut {NoStop}%
\bibitem [{\citenamefont {Steffen}\ \emph {et~al.}(2010)\citenamefont
  {Steffen}, \citenamefont {Kumar}, \citenamefont {DiVincenzo}, \citenamefont
  {Rozen}, \citenamefont {Keefe}, \citenamefont {Rothwell},\ and\ \citenamefont
  {Ketchen}}]{Steffen2010}%
  \BibitemOpen
  \bibfield  {author} {\bibinfo {author} {\bibfnamefont {M.}~\bibnamefont
  {Steffen}}, \bibinfo {author} {\bibfnamefont {S.}~\bibnamefont {Kumar}},
  \bibinfo {author} {\bibfnamefont {D.~P.}\ \bibnamefont {DiVincenzo}},
  \bibinfo {author} {\bibfnamefont {J.~R.}\ \bibnamefont {Rozen}}, \bibinfo
  {author} {\bibfnamefont {G.~A.}\ \bibnamefont {Keefe}}, \bibinfo {author}
  {\bibfnamefont {M.~B.}\ \bibnamefont {Rothwell}}, \ and\ \bibinfo {author}
  {\bibfnamefont {M.~B.}\ \bibnamefont {Ketchen}},\ }\href
  {http://link.aps.org/doi/10.1103/PhysRevLett.105.100502} {\bibfield
  {journal} {\bibinfo  {journal} {Phys. Rev. Lett.}\ }\textbf {\bibinfo
  {volume} {105}},\ \bibinfo {pages} {100502} (\bibinfo {year}
  {2010})}\BibitemShut {NoStop}%
\bibitem [{\citenamefont {Filipp}\ \emph {et~al.}(2009)\citenamefont {Filipp},
  \citenamefont {Maurer}, \citenamefont {Leek}, \citenamefont {Baur},
  \citenamefont {Bianchetti}, \citenamefont {Fink}, \citenamefont {Goppl},
  \citenamefont {Steffen}, \citenamefont {Gambetta}, \citenamefont {Blais},\
  and\ \citenamefont {Wallraff}}]{filipp_joint_2009}%
  \BibitemOpen
  \bibfield  {author} {\bibinfo {author} {\bibfnamefont {S.}~\bibnamefont
  {Filipp}}, \bibinfo {author} {\bibfnamefont {P.}~\bibnamefont {Maurer}},
  \bibinfo {author} {\bibfnamefont {P.~J.}\ \bibnamefont {Leek}}, \bibinfo
  {author} {\bibfnamefont {M.}~\bibnamefont {Baur}}, \bibinfo {author}
  {\bibfnamefont {R.}~\bibnamefont {Bianchetti}}, \bibinfo {author}
  {\bibfnamefont {J.~M.}\ \bibnamefont {Fink}}, \bibinfo {author}
  {\bibfnamefont {M.}~\bibnamefont {Goppl}}, \bibinfo {author} {\bibfnamefont
  {L.}~\bibnamefont {Steffen}}, \bibinfo {author} {\bibfnamefont {J.~M.}\
  \bibnamefont {Gambetta}}, \bibinfo {author} {\bibfnamefont {A.}~\bibnamefont
  {Blais}}, \ and\ \bibinfo {author} {\bibfnamefont {A.}~\bibnamefont
  {Wallraff}},\ }\href {\doibase 10.1103/PhysRevLett.102.200402} {\bibfield
  {journal} {\bibinfo  {journal} {Phys. Rev. Lett.}\ }\textbf {\bibinfo
  {volume} {102}},\ \bibinfo {eid} {200402} (\bibinfo {year}
  {2009})}\BibitemShut {NoStop}%
\bibitem [{\citenamefont {Chow}\ \emph
  {et~al.}(2010{\natexlab{b}})\citenamefont {Chow}, \citenamefont {DiCarlo},
  \citenamefont {Gambetta}, \citenamefont {Nunnenkamp}, \citenamefont {Bishop},
  \citenamefont {Frunzio}, \citenamefont {Devoret}, \citenamefont {Girvin},\
  and\ \citenamefont {Schoelkopf}}]{Chow2010}%
  \BibitemOpen
  \bibfield  {author} {\bibinfo {author} {\bibfnamefont {J.~M.}\ \bibnamefont
  {Chow}}, \bibinfo {author} {\bibfnamefont {L.}~\bibnamefont {DiCarlo}},
  \bibinfo {author} {\bibfnamefont {J.~M.}\ \bibnamefont {Gambetta}}, \bibinfo
  {author} {\bibfnamefont {A.}~\bibnamefont {Nunnenkamp}}, \bibinfo {author}
  {\bibfnamefont {L.~S.}\ \bibnamefont {Bishop}}, \bibinfo {author}
  {\bibfnamefont {L.}~\bibnamefont {Frunzio}}, \bibinfo {author} {\bibfnamefont
  {M.~H.}\ \bibnamefont {Devoret}}, \bibinfo {author} {\bibfnamefont {S.~M.}\
  \bibnamefont {Girvin}}, \ and\ \bibinfo {author} {\bibfnamefont {R.~J.}\
  \bibnamefont {Schoelkopf}},\ }\href
  {http://link.aps.org/doi/10.1103/PhysRevA.81.062325} {\bibfield  {journal}
  {\bibinfo  {journal} {Phys. Rev. A}\ }\textbf {\bibinfo {volume} {81}},\
  \bibinfo {pages} {062325} (\bibinfo {year} {2010}{\natexlab{b}})}\BibitemShut
  {NoStop}%
\bibitem [{\citenamefont {Motzoi}\ \emph {et~al.}(2009)\citenamefont {Motzoi},
  \citenamefont {Gambetta}, \citenamefont {Rebentrost},\ and\ \citenamefont
  {Wilhelm}}]{Motzoi:2009fx}%
  \BibitemOpen
  \bibfield  {author} {\bibinfo {author} {\bibfnamefont {F.}~\bibnamefont
  {Motzoi}}, \bibinfo {author} {\bibfnamefont {J.~M.}\ \bibnamefont
  {Gambetta}}, \bibinfo {author} {\bibfnamefont {P.}~\bibnamefont
  {Rebentrost}}, \ and\ \bibinfo {author} {\bibfnamefont {F.~K.}\ \bibnamefont
  {Wilhelm}},\ }\href {\doibase 10.1103/PhysRevLett.103.110501} {\bibfield
  {journal} {\bibinfo  {journal} {Physical Review Letters}\ }\textbf {\bibinfo
  {volume} {103}},\ \bibinfo {pages} {110501} (\bibinfo {year}
  {2009})}\BibitemShut {NoStop}%
\bibitem [{\citenamefont {Wenner}\ \emph {et~al.}(2011)\citenamefont {Wenner},
  \citenamefont {Neeley}, \citenamefont {Bialczak}, \citenamefont {Lenander},
  \citenamefont {Lucero}, \citenamefont {O’Connell}, \citenamefont {Sank},
  \citenamefont {Wang}, \citenamefont {Weides}, \citenamefont {Cleland},\ and\
  \citenamefont {Martinis}}]{Wenner2011}%
  \BibitemOpen
  \bibfield  {author} {\bibinfo {author} {\bibfnamefont {J.}~\bibnamefont
  {Wenner}}, \bibinfo {author} {\bibfnamefont {M.}~\bibnamefont {Neeley}},
  \bibinfo {author} {\bibfnamefont {R.~C.}\ \bibnamefont {Bialczak}}, \bibinfo
  {author} {\bibfnamefont {M.}~\bibnamefont {Lenander}}, \bibinfo {author}
  {\bibfnamefont {E.}~\bibnamefont {Lucero}}, \bibinfo {author} {\bibfnamefont
  {A.~D.}\ \bibnamefont {O’Connell}}, \bibinfo {author} {\bibfnamefont
  {D.}~\bibnamefont {Sank}}, \bibinfo {author} {\bibfnamefont {H.}~\bibnamefont
  {Wang}}, \bibinfo {author} {\bibfnamefont {M.}~\bibnamefont {Weides}},
  \bibinfo {author} {\bibfnamefont {A.~N.}\ \bibnamefont {Cleland}}, \ and\
  \bibinfo {author} {\bibfnamefont {J.~M.}\ \bibnamefont {Martinis}},\ }\href
  {http://iopscience.iop.org/0953-2048/24/6/065001} {\bibfield  {journal}
  {\bibinfo  {journal} {Supercond. Sci. Technol.}\ }\textbf {\bibinfo {volume}
  {24}},\ \bibinfo {pages} {065001} (\bibinfo {year} {2011})}\BibitemShut
  {NoStop}%
\bibitem [{\citenamefont {Nielsen}\ and\ \citenamefont
  {Chuang}(2000)}]{nielsen_chuang_2000}%
  \BibitemOpen
  \bibfield  {author} {\bibinfo {author} {\bibfnamefont {M.~A.}\ \bibnamefont
  {Nielsen}}\ and\ \bibinfo {author} {\bibfnamefont {I.~L.}\ \bibnamefont
  {Chuang}},\ }\href {\doibase 10.2277/0521635039} {\emph {\bibinfo {title}
  {Quantum Computation and Quantum Information}}}\ (\bibinfo  {publisher}
  {Cambridge University Press, Cambridge},\ \bibinfo {year} {2000})\BibitemShut
  {NoStop}%
\bibitem [{\citenamefont {Nielsen}(2002)}]{nielsen_gatefid_2002}%
  \BibitemOpen
  \bibfield  {author} {\bibinfo {author} {\bibfnamefont {M.~A.}\ \bibnamefont
  {Nielsen}},\ }\href {\doibase 10.1016/S0375-9601(02)01272-0} {\bibfield
  {journal} {\bibinfo  {journal} {Phys. Lett. A}\ }\textbf {\bibinfo {volume}
  {303}},\ \bibinfo {pages} {249} (\bibinfo {year} {2002})}\BibitemShut
  {NoStop}%
\end{thebibliography}
\end{document}